# Broadband Sample Holder for Microwave Spectroscopy of Superconducting Qubits

A.S. Averkin[1], A. Karpov[1], K. Shulga[1,2], E. Glushkov[1], N. Abramov[1], U. Huebner[3], E. Il'ichev[2,3], and A.V. Ustinov[1,2,4]

[1]*National University of Science and Technology (MISIS), Leninskiy prosp. 4, Moscow, 119049, Russia*

[2] *Russian Quantum Center (RQC), 100 Novaya St., Skolkovo, Moscow region, 143025, Russia*

[3] *Leibniz Institute of Photonic Technology (IPHT), PO Box 100239, D-07702 Jena, Germany*

[4] *Physikalisches Institut, Karlsruhe Institute of Technology (KIT), D-76131 Karlsruhe, Germany*

We present a practical design and implementation of a broadband sample holder suitable for microwave experiments with superconducting integrated circuits at millikelvin temperatures. Proposed design can be easily integrated in standard dilution cryostats, has flat pass band response in a frequency range from 0 to 32 GHz, allowing the RF testing of the samples with substrate size up to 4x4 mm. The parasitic higher modes interference in the holder structure is analyzed and prevented via design considerations. The developed setup can be used for characterization of superconducting parametric amplifiers, bolometers and qubits. We tested the designed sample holder by characterizing of a superconducting flux qubit at 20 mK temperature.

## I. INTRODUCTION

A frequency-dependent microwave characterization of active superconducting devices, such as Josephson qubits, represents a substantial challenge, due to the parasitic resonances and calibration problems in the circuits connecting the device located at millikelvin temperatures to the standard RF test equipment. For the RF testing in a 4 -100 Kelvin temperature range, a number of the cryogenic test setups and probe stations has been developed for the frequencies up to 200 GHz [1,2]. The RF testing in millikelvin temperature range presents an additional difficulty due to a particular complexity of the cryostat and its limited cooling power. Additionally, for characterization of the superconducting RF electronics containing Josephson junctions, one has to suppress the external magnetic fields, which in turn requires magnetic shields limiting the access to the electronic components near the tested sample.

Recently, a number of experiments reported new setups for the RF testing of superconducting electronic circuits at millikelvin temperatures [3,4]. Sample holder with no parasitic resonances for frequency band from 0 to 8 GHz was developed by Hornibrook *et al.*

[3] and another one by Chow *et al.* [4] for frequency band between 0 and 15 GHz. Normally, a cryogenic sample holder serves to establish a RF connection of the superconducting device to a coaxial cable (or a waveguide) via standard RF connectors. A low-loss connection is needed for performing the RF reflectometry of a sample using a vector network analyzer (VNA), and simplifies calibration.

In general, the reflectometry experiment presumes testing of a device embedded in a circuit with single mode transmission lines, as the concept of the reflection and transmission coefficients makes sense only in the case of the single mode propagation. Also, if a higher order mode wave can propagate inside the sample holder, or in one of its section, the unexpected parasitic resonances in the circuit may arise in the areas where the energy of the higher mode is confined. If the conditions for higher mode propagation are met, the excitation of higher mode may occur at the random small irregularities or asymmetries in the circuit, resulting from the manufacturing errors. This leads to the difficulty of analyzing data and to the problems in calibrating the test circuit. A similar interference happens when the probing signal bypasses the device through a parasitic channel opened through a higher order mode. For example, such parallel channel may induce unexpected losses and to reduce the quality factor of the CPW resonators [3].

The superconducting coplanar waveguide (CPW) is a solution of choice for designing a variety of superconducting integrated circuits, like Josephson parametric amplifiers [5-9], bolometers [10-12], and microwave circuits with superconducting qubits [13-17]. Superconducting CPWs are frequently fabricated using Nb, NbN, or Al metallization layers deposited on a silicon or sapphire substrate. The CPW line based circuits seems to be free from higher modes and related problems. The first upper mode may propagate in CPW, when the phase length of the perimeter of the central line is close to a half of the wavelength in the free space. Because the width of the central strip of the superconducting CPW may be made as small as few micrometers, the frequency limit of a single mode regime may be shifted far in the THz range, well above the device operation band. Nevertheless, the CPW is not completely problem-free. As the radiation in the CPW is not completely isolated from the surrounding space, it may couple to the waveguide modes inside of the sample housing, for example in the space over the CPW channels. Even a weak coupling to the higher mode can introduce undesirable losses or to alter the circuit spectral response. For developing broadband CPW-based circuits, it is important to avoid any excitations of the waveguide modes in the space above the circuit under study.

In this work, we describe a practical design of a sample holder for the RF characterization of superconducting circuits at millikelvin temperatures in the frequency range from 0 to 32 GHz. First, we describe details of the sample holder layout and discuss the prevention of higher mode excitation and interference. Then, we verify the frequency limits of the setup by numerical simulation and present experimental tests, including characterization of the Josephson flux qubit sample at 20 mK temperature.

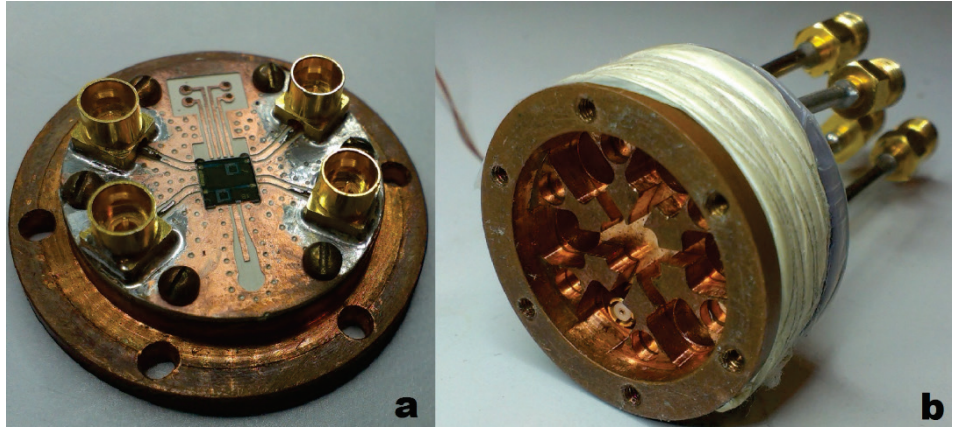

Figure 1. The bottom (a) and the top (b) parts of the developed sample holder. The top part stays attached to the cryostat. The bottom part holds a sample mounted in a slot at the center of the board. This part can be easily replaced in the cryostat. At the board, four 50 ohm coplanar lines (CPW) connect the sample ports and the surface-mounted SMP connectors. Note the via-holes with metallization preventing signal leakage and parasitic coupling inside of the holder board. The narrow channels in the top part (b) of the sample holder are reducing the space over the CPW lines, thus preventing excitation and propagation of waveguide modes above the board, and also damp parasitic couplings and cross-talks in the circuit.

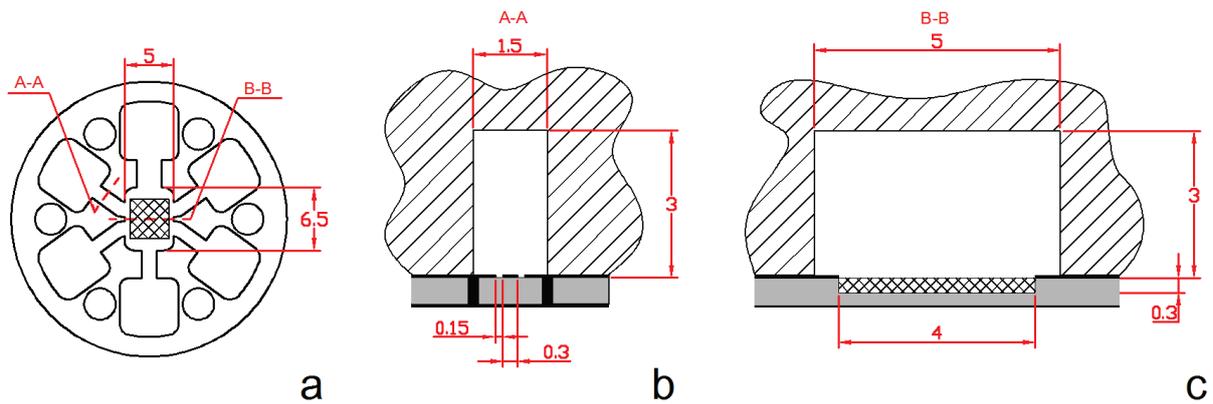

Figure 2. The sample holder design details. All dimensions are in millimeters. (a) A bottom view of the top part of the sample holder (shown in Fig.1b). The 4 X 4 mm chip location is marked with a cross-hatched area. (b) A vertical cross-section of the 1.5 mm wide channel over the CPW connecting line. The PCB board is marked by grey color, and the metallization of CPW and of the via-holes are in black. (c) Vertical cross section of the central chamber with the PCB (grey) and the substrate with a superconducting circuit (cross hatch).

## II. SAMPLE HOLDER DESIGN

We designed a cryogenic sample holder for the RF tests of superconducting electronic circuits at millikelvin temperatures. The holder consists of a copper base plate with a test board, four on-board RF connectors (Fig. 1a), and the holder cover (Fig. 1b). The holder cover normally remains attached to the cryostat. The base plate is holding a sample under test, and can be easily removed and reconnected, when needed.

The use of standard on-board SMP connectors is simplifying integration of the holder in the cryogenic test setup, allowing for a simple plug-in of the holder. The SMP is a 0-40 GHz "clickable" RF connector. This feature is particularly useful for a rapid exchange of the samples in the cryostat using a set of replaceable identical base plates. The compact arrangement of the test circuit with connectors, fitting in the 40 mm diameter, is easing the embedding of the holder in a cylindrical magnetic cryoperm shield or an external solenoid. The transmission lines connecting the sample under test and RF connectors are made as CPWs matching contact pads in the superconducting circuit. A broad slit for winding the superconducting magnetic coil is located on the external side surface of the top part of the holder (Fig. 1b). The coil is often used, e.g., in experiments with superconducting qubits. The test unit CPW has the characteristic impedance of 50 Ohm. The board material is 0.5 mm thick dielectric AD1000 with dielectric constant $\varepsilon_r$ =10.6. The via-holes spaced by 1 mm are placed on both sides of the CPW in order to prevent excitation of the unwanted substrate modes.

Miniaturization of the test circuit helps pushing the higher mode frequencies above the operation band. The limit of holder shrinking is set by the size of the tested substrate with the superconducting circuit and RF connectors. In our qubit experiments, a typical size of the substrate is 4 X 4 mm$^2$. In order to accommodate the sample, a chamber with a footprint of 5 X 6.5 mm$^2$ is left above the circuit substrate in the middle of the top part of the holder. The sample is thus coupled to the RF connectors by the CPW lines running in the narrow channels (Figs. 1, 2). The channel width and height are set small enough to prevent propagation of the waveguide modes in the frequency band of interest. The cross-section area of the waveguide mode propagation is limited by the channel walls, by the via-holes in the printed circuit board (PCB), and by the bottom metallization of the PCB (Fig. 2).

The narrow channel above the PCB (Fig. 2b) may be considered as a rectangular waveguide. The cut-off frequency $f_{Cwg}$ for the first mode of a rectangular waveguide is:

$$f_{Cwg} = \frac{c}{2\,a\,\sqrt{\varepsilon_{eff}}} \qquad (1)$$

where, $a$ is the width of the waveguide, $c$ is the speed of the light in free space and $\varepsilon_{eff}$ is the effective dielectric constant in the waveguide. At the first waveguide mode, when a thin dielectric layer of the CPW board is located along the narrow wall, the effective dielectric constant is close to unity [18]. Putting the channel size at 1.5 X 3 mm allows avoiding propagation of the waveguide modes in the channel up to 50 GHz. The height of the channel is sufficient to decrease the parasitic capacitance between the CPW elements and the channel walls.

As the first order approximation, the cavity over the chip may be considered as a rectangular waveguide resonator. The lowest resonance frequency mode of the cavity is given by the expression:

$$f_{Cav} = \frac{c\,\sqrt{d^2+l^2}}{2\,d\,l} \qquad (2)$$

Where $c$ is the speed of light in the free space, $d$ and $l$ the width and lengths of the cavity. For a cavity size of $d$ = 6.5 mm by $l$ = 5 mm, the frequency $f_{Cav}$ = 37.9 GHz, which is well above the targeted 30 GHz limit. The effective volume of the resonator is increased due to the channel openings at the edges of the chamber (Fig. 2), and the resonance frequency is somewhat lower than estimated with simplified approach taken in Eq.(2). A more detailed numerical analysis of the waveguide mode resonance is presented below.

A numerical model of electrodynamics of the holder structure was prepared with ANSYS HFSS program package. In calculations, we used a detailed 3D drawing of the holder with printed circuit and connectors, as shown in Fig. 2. The channel going upward from the central chamber in Fig. 2 is 2.5 mm wide; it is used for 4 DC lines. All remaining channels used for CPW lines are 1.5 mm wide. The chamber length is fixed to be 5 mm, and its width is varied in simulations. The circuit operates without unexpected resonances, as it should be with the mono-mode coplanar waveguide lines, up to 32 GHz (Fig. 3). The first resonance feature appears in the transmission coefficient versus frequency dependence ($|S21(f)|$) above 32 GHz and depends on the central chamber dimensions. As the different sample size may require adjustment of the cavity size, we are giving calculated resonance frequencies for the 5X5 mm, 5X6.5 mm, and 5X8mm in Table 1. In our experiment, the chamber size is chosen to be 5X6.5 mm.

Table I. Calculated waveguide mode resonance frequency versus size of the space over the sample.

| Cavity size [mm] | Resonance frequency [GHz] |
|---|---|
| 5 X 5.0 | 35.05 |
| 5 X 6.5 | 32.30 |
| 5 X 8.0 | 30.21 |

The developed sample holder (with cavity size 6.5 X 5 mm) operates in mono-mode regime up to 32 GHz (Figs. 3, 4). The HFSS simulated electrical field distribution in the space above the sample for the first resonance frequency $f_1$ of 32.3 GHz is presented in Fig. 4a. Once can see that the resonating electromagnetic field is located in the central cavity, over the substrate with the circuit under test. This feature may stay unnoticed if the device under the test has low transmission losses at the resonance frequency $f_1$. In case of a low transmission through the device under the test, the parasitic resonance is more visible. For example, with a break in the central line in the middle of the sample, the transmission is strongly affected by the resonance at $f_1$ (Fig. 4b). In Fig.4b the transmission S21(f) and reflection S11(f) coefficients are calculated for a broken central line with a gap, acting as a small capacitor. The interaction between the CPW line and the waveguide resonator, formed between the board and the cover, is strong enough to create a significant parasitic coupling at the resonance frequency. Note, that at 32.3 GHz the waveguide mode is confined to the space over the sample, and is not propagating over the narrow channels through the inner space of the holder.

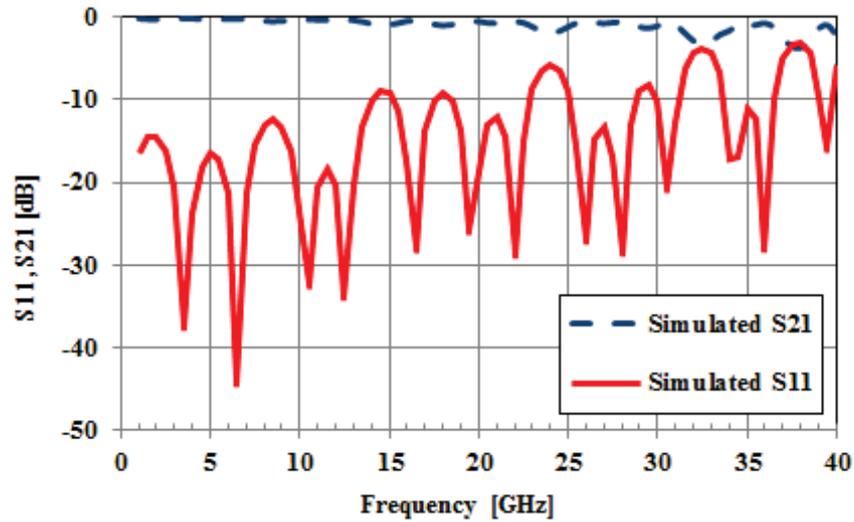

Figure 3. Calculated transmission S21(f) (dashed line) and reflection S11(f) (solid line) in the sample holder with a section of a 50 Ohm coplanar line in place of the superconducting circuit. The sample holder is well matched with coaxial waveguide up to 32 GHz. The details of the resonance at 32.3 GHz are presented in Fig. 4.

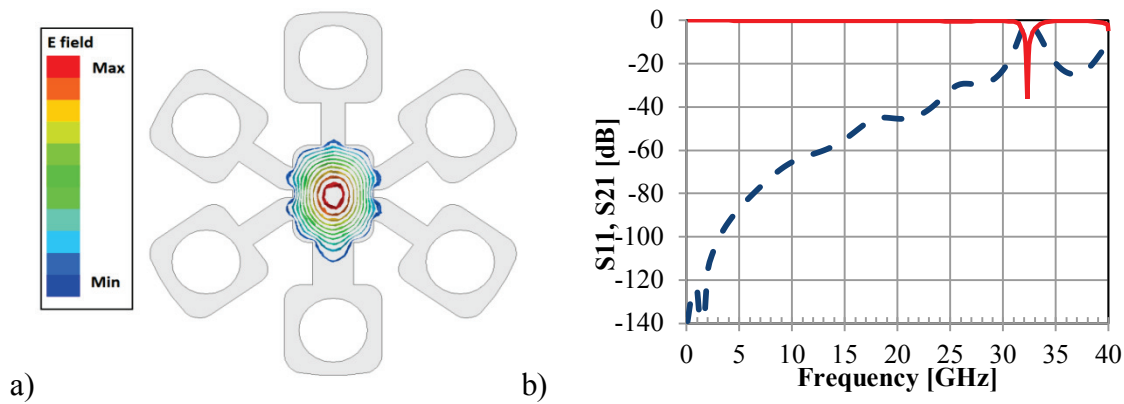

a) b)

Figure 4. The developed sample holder operates in a mono-mode regime up to 32 GHz. The first parasitic cavity mode resonance occurs at $f_1$ = 32.3 GHz. In a case of a low transmission through the device under the test, the effect of the resonance is more visible. For example, if a capacitive break is introduced in the central line of a 50 Ohm CPW in the middle of the board, the transmission is strongly affected by the resonance at 32.3 GHz. a) Distribution of the amplitude of the electric field at 33.2 GHz in the space over the board. Note, at 32.3 GHz the energy is confined to the space over the sample. b) Signal propagation through the holder with a capacitive break in the CPW line in the middle of the circuit. Here transmission S21 is given with dashed line and reflection S11 – with solid line. High transmission at 32.3 GHz is due to the waveguide mode excited in the space around the sample.

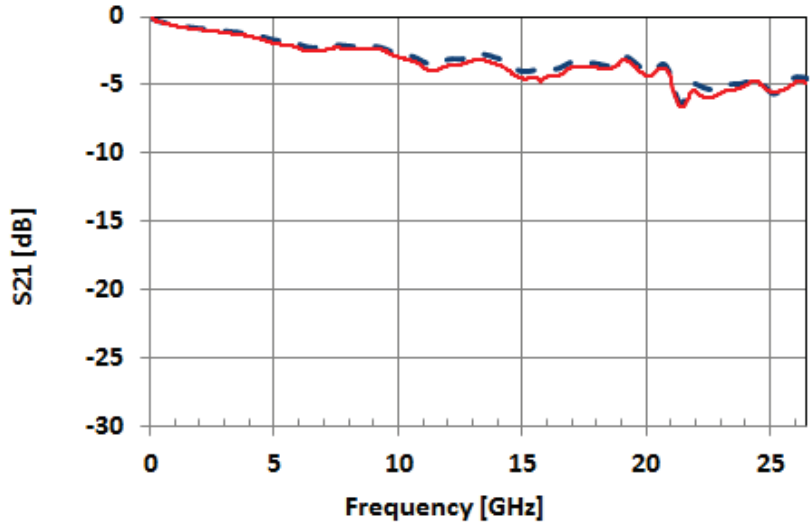

Fig.5. Test measurement of the sample holder at room temperature. Transmission through the sample holder with a 50 Ohm line section in place of the sample is measured with (solid line) and without (dashed line) the cover. Because the waveguide modes are excited in the space under the cover, a difference would indicate the presence of the parasitic resonances. These two measurements gave nearly identical results, confirming that the sample holder is free of parasitic resonances in 0 – 27 GHz frequency band. The 5 dB loss in connectors and cables at 25 GHz is acceptable for our application.

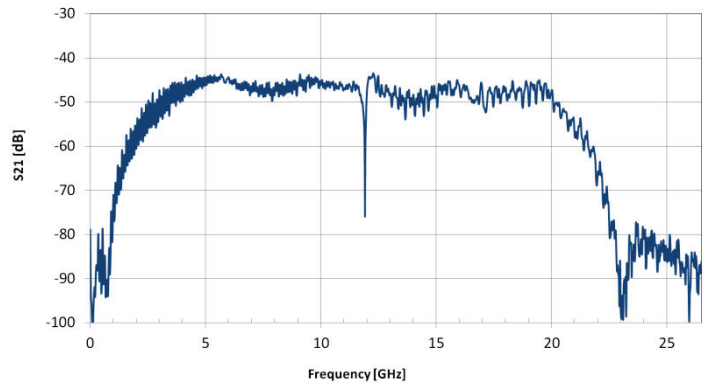

Figure 6. An example of use of the developed RF sample holder. The measured transmission is the RF response of the readout circuit of the Josephson flux qubit. The flux qubit is weakly coupled to a coplanar resonator with the resonance frequency of about 12 GHz. The measurement is performed at 20 mK temperature. The test bandwidth is limited to 3 – 20 GHz by the band of the cryogenic low noise amplifier. Note a uniform transmission and the absence of the parasitic resonances in the frequency range of interest.

## III. EXPERIMENT

In experiment we checked first the sample holder for parasitic resonances in a simplest configuration, with a section of 50 Ohm coplanar line instead of the sample. Installing a section of 50 Ohm coplanar line in place of the sample, we measured transmission without cover, and then with cover attached. Because the waveguide modes are excited in the space under the cover, a difference would indicate the presence of related parasitic resonances. The measured transmission through the holder does not change when the cover is in place, and no visible marks of parasitic resonances in |S21(f)| in the 0.01 – 27 GHz band (Fig. 5), confirming the predictions of our numerical simulations. The measured loss is mainly related to the cables and connectors used in the test. It is rising from 0 dB at low frequencies, up to 5 dB at 25 GHz. This level of loss is acceptable for cryogenic setup, where 10 – 30 dB attenuators are usually used for suppression of the room temperature thermal noise, coming through the connecting cables. In the measurements, we used a 0.01-26.5 GHz VNA Agilent PNA-X.

We tested the developed RF sample holder to characterize a Josephson flux qubit. In the tested device, the qubit is embedded in a CPW resonator, which in turn is weakly capacitively coupled to a control CPW line on the chip [19]. The circuit response is observed to be a narrow resonance absorption peak in S21(f) around 12 GHz (Fig. 6). The measurements where performed at 20 mK temperature in a dilution cryostat. In these measurements, our cryogenic 3-20 GHz low noise amplifier is limiting the band of observation. Note a uniform transmission in the band of interest and the absence of the parasitic resonances in the measured data.

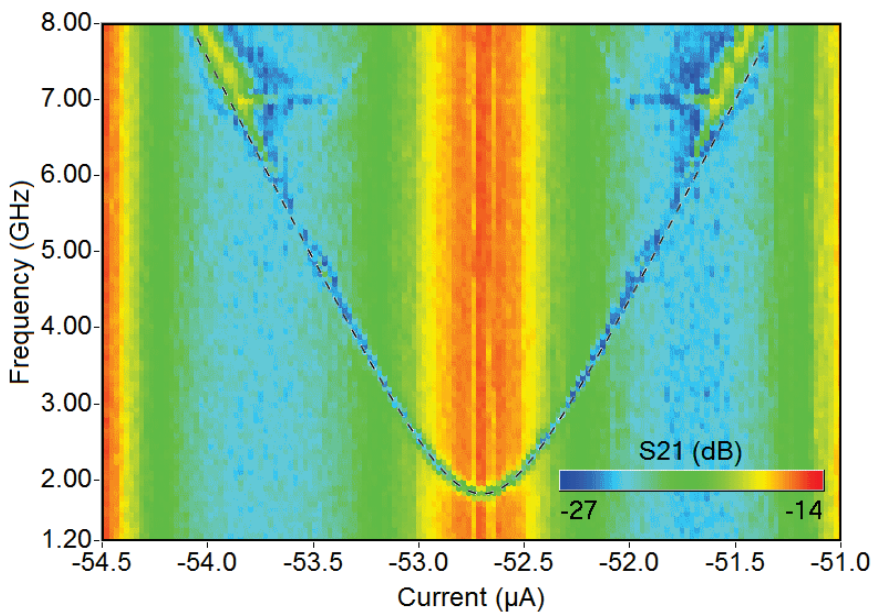

Fig. 7. An example of the measured energy spectrum (|0> - |1> transition) of the flux qubit as a function of magnetic field (*i.e.* function of the current applied to magnetic bias coil). The superconducting loop of the qubit is weakly coupled to λ/4 readout resonator. The minimal level splitting (Δ≈1.8GHz) and position of degeneracy point of the qubit are obtained from this data.

Finally, we used the developed sample holder to perform full-band spectroscopy of a superconducting flux qubit (Fig. 7). In this case, the readout microwave tone was applied via a coplanar line connected to seven λ/4 resonators with different frequencies. A flux qubit was composed of a superconducting ring with three Josephson junctions and inductively coupled to the corresponding resonator [19, 20]. In addition to the readout tone, another spectroscopy probe microwave tone was applied at a frequency swept in a wide range. A dispersive shift of the resonator occurred when the qubit got excited between its ground and excited states, and detected with vector network analyzer measuring amplitude and phase of the microwave readout signal passing through the sample, as a function of the probing signal power and frequency. The readout signal transmitted through the sample was amplified using a cryogenic amplifier at 4 K and then a chain of room-temperature amplifiers. We used attenuators at the input and isolators in the output line of the chip in order to isolate qubit from thermal noise sources and signal reflections. The energy spectrum of the qubit was controlled by the external magnetic field, produced by a superconducting coil.

The measured spectrum of a flux qubit (Fig. 7) is free of parasitic resonances and gives a typical hyperbolic-like dependence of the qubit frequency versus magnetic flux (proportional to the current applied to bias coil). The dependence of the qubit frequency versus $\Delta$ and flux detuning $\varepsilon$ is given by equation $\omega_q = \sqrt{\Delta^2 + \varepsilon^2}$ and allows for determining the minimal level splitting for the qubit symmetry point of about $\Delta \approx 1.8 \text{GHz}$. The thickening of the qubit hyperbolic dependence near 7 GHz can be associated with a resonance in the on-chip qubit readout circuit.

## IV. SUMMARY

A broadband cryogenic sample holder is developed for characterization of superconducting circuits at mK temperatures. The designed holder may house samples of a typical size of 4 X 4 mm$^2$, and allows an external RF connection to the sample via four coplanar waveguide lines. The possibility of higher modes interference in the holder structure is analyzed and prevented. A particular attention is paid to the possibility of excitation of waveguide modes propagating (or trapped) in the space over the sample. The numerical simulation of the electrodynamics of the holder is used to confirm the possibility of the waveguide modes excitation in inner cavities of the holder, and their role in forming parasitic resonances. The developed sample holder is free of parasitic resonances in the frequency band from 0 to 32 GHz. Adding two extra RF connectors with coplanar waveguide lines to this particular sample holder, thus increasing the total number of test lines to six, appears straightforward. The developed holder has been used for detecting the dispersive response and measuring the spectrum of a Josephson flux qubit at 20 mK temperature. The designed setting can be useful for RF characterization of superconducting parametric amplifiers, bolometers, and qubits.

## AKNOWLEDGMENTS

This work was supported in part by the Ministry of Education and Science of the Russian Federation under contract no. 11.G34.31.0062. EI and UH acknowledge the European Community's Seventh Framework Programme (FP7/2007-2013) under Grant No. 270843 (iQIT).